\documentclass[journal,twocolumn]{IEEEtran}
\usepackage{ifpdf}
\usepackage{cite}
\usepackage[cmex10]{amsmath}
\usepackage{algorithmic}
\usepackage[tight,footnotesize]{subfigure}
\usepackage{fixltx2e}
\usepackage[pdftex]{graphicx}

\begin{document}
\title{A Simple Cooperative Transmission Protocol for Energy-Efficient Broadcasting Over Multi-Hop Wireless Networks}
\author{\IEEEauthorblockN{Aravind Kailas,}
\and \IEEEauthorblockN{Lakshmi Thanayankizil,} \and
\IEEEauthorblockN{and~Mary Ann Ingram\\}
\thanks{Mary Ann Ingram, Aravind Kailas and Lakshmi Thanayankizil are with
the School of Electrical and Computer Engineering, Georgia Institute
of Technology, Altanta, GA 30332-0250, USA. The authors gratefully acknowledge support for this research from the National Science Foundation under CNS-0721269 and CNS-0721296. }}
\maketitle

\begin{abstract}
This paper analyzes a broadcasting technique for wireless multi-hop sensor networks that uses a form of cooperative diversity called opportunistic large arrays (OLAs). We propose a method for autonomous scheduling of the nodes, which limits the nodes that relay and saves as much as 32\% of the transmit energy compared to other broadcast approaches, without requiring Global Positioning System (GPS), individual node addressing, or inter-node interaction. This energy-saving is a result of cross-layer interaction, in the sense that the Medium Access Control (MAC) and routing functions are partially executed in the Physical (PHY) layer. Our proposed method is called OLA with a transmission threshold (OLA-T), where a node compares its received power to a threshold to decide if it should forward. We also investigate OLA with variable threshold (OLA-VT), which optimizes the thresholds as a function of level. OLA-T and OLA-VT are compared with OLA broadcasting without a transmission threshold, each in their minimum energy configuration, using an analytical method under the orthogonal and continuum assumptions. The trade-off between the number of OLA levels (or hops) required to achieve successful network broadcast and transmission energy saved is investigated. The results based on the analytical assumptions are confirmed with Monte Carlo simulations.
\end{abstract}

\IEEEpeerreviewmaketitle


\IEEEpeerreviewmaketitle

\section{Introduction}
\IEEEPARstart{C}{ooperative} transmission for Wireless Sensor Networks (WSNs) is a relay technique where multiple, spatially separated radios cooperate to transmit the same message so that the receiver can derive diversity gain from the multiple transmissions \cite{sendonaris}, \cite{laneman}. Because of the diversity gain, the transmitters can dramatically lower their transmit powers and save energy without sacrificing reliability.  This paper proposes a ``transmission threshold" method for autonomous node scheduling as an extension to the existing Opportunistic Large Array (OLA) cooperative transmission broadcast protocol \cite{hong}. We will refer to the existing method as `Basic OLA' in this paper.

In a multi-hop ad hoc network, broadcasting is a significant operation to support numerous applications. For example, broadcasting is used in ``Hello" messages and route discovery in ad hoc networks and queries in WSNs. Flooding, one of the earliest broadcast protocols for multi-hop transmissions, where all nodes relay the received message, is energy inefficient and unreliable, as it leads to severe contention, collision and redundancy, a situation referred to as \textit{broadcast storm} \cite{Ni}.

Many broadcast strategies have been proposed to avoid the broadcast storm. One energy efficient type of broadcast strategy is the broadcast tree. A popular example is the Broadcast Incremental Power (BIP) algorithm, which  was proposed and analyzed in  \cite{wieselthier}.  In a broadcast tree, the relay nodes are carefully chosen so that, ideally, each node receives the message the first time from exactly one relay.  While this approach reduces considerably the number of collisions in a broadcast, it is fundamentally inconsistent with cooperative transmission. Furthermore, the identification of relay nodes in a broadcast tree requires significant network overhead, especially for mobile networks.

Our proposed protocol  does not require that a node knows its geographical location, so we mention a few other broadcasting protocols that also do not require this information. The Border Node Retransmission Based Probabilistic Protocol privileges the retransmission by nodes located at the radio border of the transmitter \cite{cartigny}. Border nodes are  identified through single hop exchanges of ``Hello" messages and hence this scheme requires no location or signal strength information. In another work by Cartigny et al. \cite{cartigny1}, the authors considered a Relative Neighbor Graph (RNG) relay subset protocol \cite{cartigny1}, where only a subset of nodes relay the message from the source. Pairs of nodes are assumed to be able to evaluate their relative distance with integration of a positioning system or a signal strength measure. Since both \cite{cartigny} and \cite{cartigny1} require neighbor information, these protocols will not scale well with node density.

An OLA is a collection of nodes that transmit the same message at approximately the same time. They do this without coordination between each other, but they naturally fire at approximately the same time in response to energy received from a single source or another OLA \cite{ola}. Carrier sensing must be disabled to permit OLA transmission to take place. Because all the transmissions within an OLA are repeats of the same message, the signal received from an OLA has the same model as a multipath channel. Small time offsets (because of different distances and computation times) \cite{wei}, and small frequency offsets (because each node has a different oscillator frequency) appear as excess delays and Doppler shifts, respectively. As long as the receiver, such as a RAKE receiver, can tolerate the effective delay and Doppler spreads of the received signal, decoding should proceed normally.

The primary benefits of an OLA transmission are its spatial diversity gain and its lack of network overhead. Even though many nodes may participate in an OLA transmission, energy can still be saved because all nodes can reduce their transmit powers dramatically and large fade margins are not needed. Even in non-fading channels, the array gain in an OLA transmission may be desirable for applications where there is a low maximum power constraint, resulting, for example from severe cost or heat restrictions. Further, in \cite{ola}, the Basic OLA algorithm was shown to yield an energy savings of about 5 dB compared to the BIP algorithm.

Other OLA works include the following. In the centralized broadcasting scheme \textit{Accumulative Broadcast} \cite{ivana}, the order and power levels of node transmissions are chosen to minimize the total power consumption in general multi-hop networks. While the second problem can be solved by means of linear programming tools, the first problem was found to be NP complete \cite{ahluwalia}, \cite{liang}. In \cite{optimal}, an optimal \textit{trivial schedule} was found for a dense OLA network that allocated power and order of transmission according to node distance from the source.

The Dual Threshold Cooperative Broadcast (DTBC) was introduced in \cite{power} as a way to save even more energy compared to the Basic OLA broadcast, by allowing a node to join an OLA only if its received signal power is less than a given threshold. However the DTBC concept was not analyzed in \cite{power}.  Our paper analyzes OLA-Threshold (OLA-T), which is the same, but independently derived, concept as DTBC. Our paper also extends the concept to allow the thresholds to vary from OLA to OLA. OLA with variable threshold (OLA-VT) can be optimized to minimize total energy in a broadcast.  OLA-VT can also be used to control OLA sizes, thereby enabling certain other protocols, such as the OLA Concentric Routing Algorithm (OLACRA), which does upstream routing in WSNs \cite{sensorcomm}. OLA-T and OLA-VT can both be shown to be suboptimal trivial schedules \cite{optimal}, with the virtues of simple implementation and good performance.

Finally, two important features that all the proposed schemes inherit from Basic OLA is that individual nodes are not addressed and they do not need location information for routing. Lack of addressing makes the protocols scalable with node density. Not needing location knowledge for routing makes the proposed protocol suitable for applications where location information is either not available or too expensive or energy consuming to obtain or exploit.

\section{System Model}
Half-duplex nodes are assumed. For the purpose of analysis, the nodes are assumed to be distributed uniformly and randomly over a continuous area with average density $\rho$. The originating node is assumed to be a point source at the center of the given network area. We assume a node can decode and forward a message without error when its received Signal-to-Noise Ratio (SNR) is greater than or equal to a modulation-dependent threshold \cite{asympanal}. Assumption of unit noise variance transforms the SNR threshold to a received power criterion, which is denoted as the decoding threshold $\tau_{d}$. We note that the decoding threshold $\tau_{d}$ is not explicitly used in real receiver operations. A real receiver always just tries to decode a message. If no errors are detected, then it is assumed that the receiver power must have exceeded $\tau_{d}$. In contrast, the proposed transmission threshold of OLA-T would be explicitly compared to an estimate of the received SNR.

Let the source power, relay transmit power, and the relay transmit power per unit area be denoted by $P_{s}$,  $P_{r}$, and $\overline{P_{r}}=\rho P_{r}$, respectively. Following \cite{asympanal}, we assume the \textit{deterministic model}, which implies that the power received at a node is the sum of the powers from each of the node transmissions. This model implies node transmissions are orthogonal. However, because non-orthogonal transmissions also produce similarly shaped OLAs \cite{asympanal}, the basic OLA-T concept should work for them as well although the theoretical results would have to be modified. Furthermore, we show in Section \ref{sec:continuum_validity} that the deterministic result can approximated in a Rayleigh faded channel by using distributed 4-th order transmit diversity in the OLA transmissions. Again following \cite{asympanal}, we assume a continuum of nodes in the network, which means that we let the node density $\rho$ become very large ($\rho\rightarrow\infty$) while $\overline{P_{r}}$ is kept fixed. Section \ref{sec:continuum_validity} will also show that results of the continuum assumption can be approximated with a ``reasonable" node density for a network.

The path loss function in Cartesian coordinates is given by $l(x,y)=(x^{2}+y^{2})^{-1}$, where $(x,y)$ are the normalized coordinates at the receiver. As in \cite{asympanal}, distance $d$ is normalized by a reference distance, $d_{0}$. Let power $P_{0}$ be the received power at $d_{0}$. Received power from a node distance $d$ away is $P_{\textrm{rec}}= \min(\frac{P_{0}}{d^{2}},P_{0})$\cite{asympanal}. As in \cite{asympanal}, the aggregate path-loss from a circular disc of radius $r_{0}$ at an arbitrary distance $p>d_{0}$ from the source is given by 
\begin{eqnarray}\label{eq:path_loss}
f(r_{0},p)&=&\displaystyle \int_{0}^{r_{0}}\int_{0}^{2\pi}l(p-r \cos\theta,r\sin\theta)rdrd\theta,\nonumber{}\\
&=&\pi \ln \frac{p^2}{|p^2-r_{0}^2|}.
\end{eqnarray}  
Then the received power at a distance $p$ from the source, $P_{p}$ is given by $P_{p}= \overline{P_{r}}\pi \ln \frac{p^2}{|p^2-r_{0}^2|}$. We note that the normalized relay transmit power, $\overline{P_{r}}$, is actually the SNR received by a node at the reference distance away from a single relay node.

\section{Basic OLA}
\begin{figure}[htp]
\centering
\includegraphics[width = 2.7in]{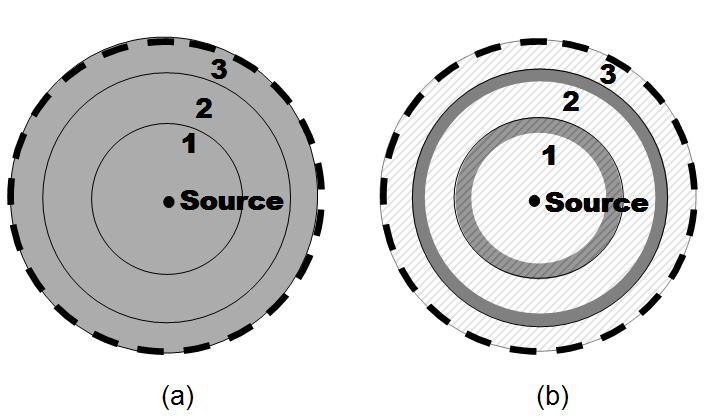}
\caption{\footnotesize \sl (a) Broadcast using Basic OLA,~(b) Broadcast using OLA with transmission threshold (OLA-T). Only nodes in the grey areas relay.}\label{fig:plot01}
\end{figure}

In a Basic OLA broadcast \cite{ola}, a node relays immediately if it can decode and if it has not relayed before. The aim is to succeed in broadcasting the message over the whole network. The source node transmits a message and the group of neighboring nodes that receive and decode the message form Decoding Level 1 ($DL_{1}$), which is the disk enclosed by the smallest circle in Fig. \ref{fig:plot01}(a). Next, each node in $DL_{1}$ transmits the message. These transmitting nodes in $DL_{1}$ constitute the first OLA. Next, nodes outside of $DL_{1}$ receive the superposition of relayed copies of the message. Nodes in this group that can decode the message constitute $DL_{2}$, which are represented as the ring
between $DL_{1}$ and the next bigger concentric circle in Fig. \ref{fig:plot01}(a). All the nodes in a decoding level form an OLA, which in turn generates the next decoding level. From \cite{asympanal}, for a fixed $\overline{P_{r}}$, the maximum value of $\tau_{d}$ such that the relayed signal will be propagated in a sustained manner by concentric OLAs satisfies 
\begin{equation}\label{eq:successcriteria}
\tau_{d}\leq \pi (\ln 2)\overline{P_{r}}.
\end{equation}
Fig. \ref{fig:plot01}(a) illustrates this phenomenon for a given network area (defined in Fig. \ref{fig:plot01} by the dashed line).

\section{OLA with Transmission Threshold (OLA-T)}\label{sec:olat}
The energy efficiency of OLAs can be improved preventing the nodes whose transmissions have a negligible effect on the formation of the next OLA from participating in the relaying. By definition, a node is near the forward boundary if it can only barely decode the message. The state of \textit{barely decoding} can be determined in practice by measuring the average length of the error vector (the distance between the received and detected points in signal space), conditioned on a successful CRC check. On the other hand, a node that receives much more power than is necessary for decoding is more likely to be near the source of the message. The OLA-T method is simply Basic OLA with the additional transmission criterion that the node's received SNR must be less than a specified \textit{transmission threshold},  $\tau_{b}$. The difference between the two thresholds is given by $\tau_{b} - \tau_{d}=\epsilon$. We will also refer to the Relative Transmission Threshold (RTT), defined as $\frac{\tau_b}{\tau_d}$. In other words, $\tau_{d}$ and $\tau_{b}$ define the range of received powers that correspond to the ``significant" boundary nodes. The concept and analysis of OLA-T are the original contributions of this paper.

Fig. \ref{fig:plot01}(b) illustrates this concept. The grey strips in Fig. \ref{fig:plot01}(b) represent OLAs within each decoding level. Unlike the approach depicted in Fig. \ref{fig:plot01}(a), the nodes that compose an OLA are only a subset of the nodes in a decoding level.

\begin{figure}[htp]
\centering
\includegraphics[width = 3.3in]{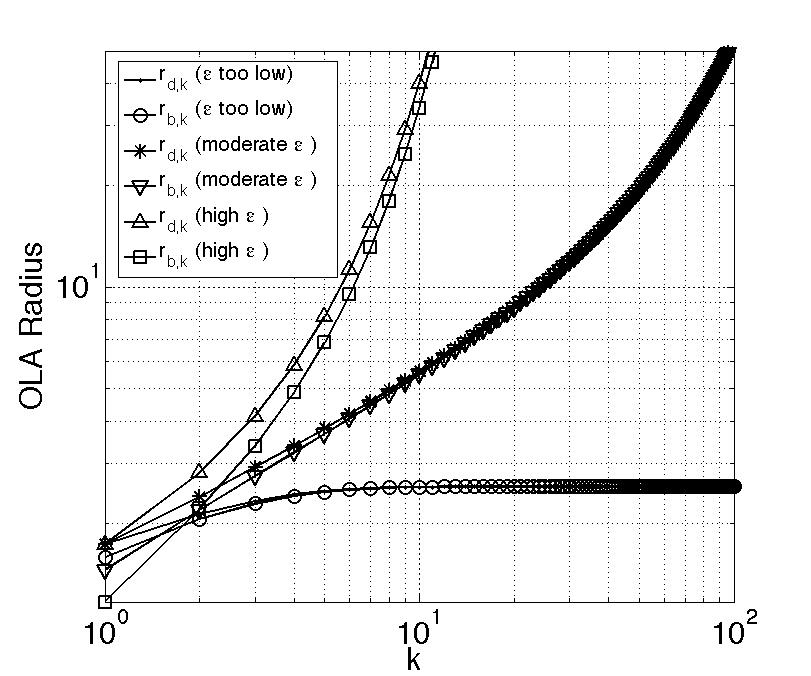}
\caption{\footnotesize \sl Outer radii, $r_{d,k}$, and the inner radii, $r_{b,k}$ versus $k$}\label{fig:ola_radii_versus_level}
\end{figure}

Before the analysis of OLA-T, it is important to point out that the transmission threshold, $\tau_{b}$, is only one of the ways to achieve energy savings. For example, it is also possible to save on energy by varying the relay transmission power, $P_{r}$, of the sensors (depending on their level) across the network. OLA-T can be thought of as an extreme quantization of variable power allocation, and therefore will not be as power efficient as an optimal continuous power allocation. However OLA-T has the advantage of essentially no network overhead, making it potentially applicable to highly mobile networks.

We note that Basic OLA transmission has been proposed for unicast transmission because of its lack of overhead \cite{biagioni}. For radios that consume substantial energy when receiving and decoding, Basic OLA might not be advantageous for unicast, since every node receives and decodes. However, in OLA-T, a node doesn't need to decode the data if it is not a relay and not the destination. If the energy spent determining that a node should relay can be made extremely small, then OLA-T might be an attractive unicast scheme. However, in this paper we consider only broadcasting and transmit energy.

We acknowledge that although OLA-T saves energy compared to Basic OLA in a single broadcast, the nodes selected by OLA-T for relaying will drain their batteries quickly, because the same nodes are always selected for a fixed source in a static network. In this case, OLA-T would cause a network partition even earlier than Basic OLA because the relays use a slightly higher transmit power. However, the opposite will be true if the source location varies randomly, or if the nodes move about randomly.  Even for a fixed source and a static network, network lifetime can be extended relative to Basic OLA by modifying OLA-T to use mutually exclusive sets of nodes on consecutive broadcasts. This new technique, which we call Alternating OLA-T (A-OLA-T) \cite{globecom08} builds on the results reported in this paper.

Finally, we note that to decode, a node in an OLA-T network receives energy from just one decoding level. Multiple levels are not ganged to form a very thick OLA as in \cite{power} nor are OLA transmissions at different times from different decoding levels combined as in \cite{ivana}. Instead, the emphasis of OLA-T is on forming thin, widely separated OLAs.

\subsection{Analysis of OLA-T Broadcast for Constant $\epsilon$}\label{sec:olata}
In this section, the OLA boundaries are determined as functions of the decoding level $k$, for the case when the transmission threshold is constant over the network. The case of variable $\tau_{b}$ is treated in Section \ref{sec:olavt}. For the constant $\epsilon$, we are able to derive the closed-form expression by slightly modifying the continuum approach in \cite{asympanal}.

Let the outer and inner boundary radii for the $k$-th OLA ring be denoted as $r_{d,k}$ and $r_{b,k}$, respectively. The boundaries can be found recursively using
\begin{equation}\label{eq:recursive1}
\overline{P_{r}}\left[f(r_{d,k},r_{j,k+1})-f(r_{b,k},r_{j,k+1})\right]=\tau_{j},~j\in\{b,d\}.
\end{equation}
Applying (\ref{eq:path_loss}) yields $\frac{\tau_{j}}{\overline{P_{r}}}=\pi\ln\frac{|r_{j,k+1}^2-r_{b,k}^2|}{|r_{j,k+1}^2-r_{d,k}^2|}$, which yields
\begin{equation}\label{eq:olaradii1}
r_{d,k}^{2}=\frac{\beta(\tau_{d})r_{d,k-1}^2 -r_{b,k-1}^2}{\beta(\tau_{d})-1},~
r_{b,k}^{2}=\frac{\beta(\tau_{b})r_{d,k-1}^2 -r_{b,k-1}^2}{\beta(\tau_{b})-1},
\end{equation}
where $\beta(\tau)=\exp\left[\tau/(\pi\overline{P_{r}})\right]$. Next, the recursive problem is cast as a matrix difference equation as follows:
\begin{eqnarray}\label{eq:difference_equation}
\left[\begin{array}{c}r_{d,k+1}^{2}\\r_{b,k+1}^{2}\end{array}\right]
&=&\left[\begin{array}{cc}\alpha(\tau_{d})+1&-\alpha(\tau_{d})\\\alpha(\tau_{b})+1&-\alpha(\tau_{b})\end{array}\right]\left[\begin{array}{c}r_{d,k}^{2}\\r_{b,k}^{2}\end{array}\right],\nonumber{}
\end{eqnarray}
where $\alpha(\tau)=\left[\beta(\tau)-1\right]^{-1}$.

Using the initial conditions, $r_{d,1}=\sqrt{\frac{P_{s}}{\tau_{d}}}$, and $r_{b,1}=\sqrt{\frac{P_{s}}{\tau_{b}}}$, the following solution is obtained by solving this first-order difference equation using Z-transforms:

\begin{equation}\label{eq:olaradii2}
r_{d,k}^{2}=\frac{\eta_{1}^{k}-\eta_{2}^{k}}{A_{1}-A_{2}},~
r_{b,k}^{2}=\frac{\zeta_{1}^{k}-\zeta_{2}^{k}}{A_{1}-A_{2}},
\end{equation}
where $A_{1}=\alpha(\tau_{d})-\alpha(\tau_{b})$, $A_{2}=1$, $A_{1}\neq1$, for $i\in\{1,2\}$,
\begin{equation}
\eta_{i}^{k}=\left\{\left[A_{i}+\alpha(\tau_{b})\right]\frac{P_{s}}{\tau_{d}}-\alpha(\tau_{d})\frac{P_{s}}{\tau_{b}}\right\}(A_{i})^{k-1}, \textrm{ and }\nonumber{}
\end{equation}
\begin{equation}
\zeta_{i}^{k}=\left\{\left[1+\alpha(\tau_{b})\right]\frac{P_{s}}{\tau_{d}}+\left[A_{i}-\alpha(\tau_{d})-1\right]\frac{P_{s}}{\tau_{b}}\right\}(A_{i})^{k-1}.\nonumber{}
\end{equation}

The radii given by (\ref{eq:olaradii2}) have been plotted in Fig.\ref{fig:ola_radii_versus_level} on a logarithmic scale, as functions of the OLA index. The low, moderate, and high values of $\epsilon$ are 0.2, 0.43, and 1.2 (in terms of RTT in dB, 0.79, 1.55, and 3.42), respectively. Where network broadcast is achieved, the radii grow in an unbounded fashion, with a rate that increases with level index, $k$. We have observed that for some values of $\epsilon$, such as $\epsilon=0.43$, the radii increase at a sub-linear rate with respect to $k$, up to a certain point and then the increases are faster than linear for all higher $k$ (that we test).

From observation of the above development and of the Basic OLA condition for broadcast in (\ref{eq:successcriteria}), we find that the variables $\tau_{d}, \tau_{b}, \epsilon$, and $P_s$ always appear divided by $\overline{P_{r}}$ ($\overline{P_{r}}$ cancels in the ratio $P_s/\tau_d$). In particular, we give $\tau_d/\overline{P_{r}}$ the name Decoding Ratio (DR), because it can be shown to be the ratio of the receiver sensitivity (i.e. minimum power for decoding at a given data rate) to the power received from a single relay at the `distance to the nearest neighbor,' $d_{nn}=1/\sqrt{\rho}$. If $\rho$ is a perfect square, then the $d_{nn}$ would be the minimum distance between the nearest neighbors if the nodes were arranged in a uniform square grid. 
\begin{figure}[htp]
\centering
\includegraphics[width = 3.3in]{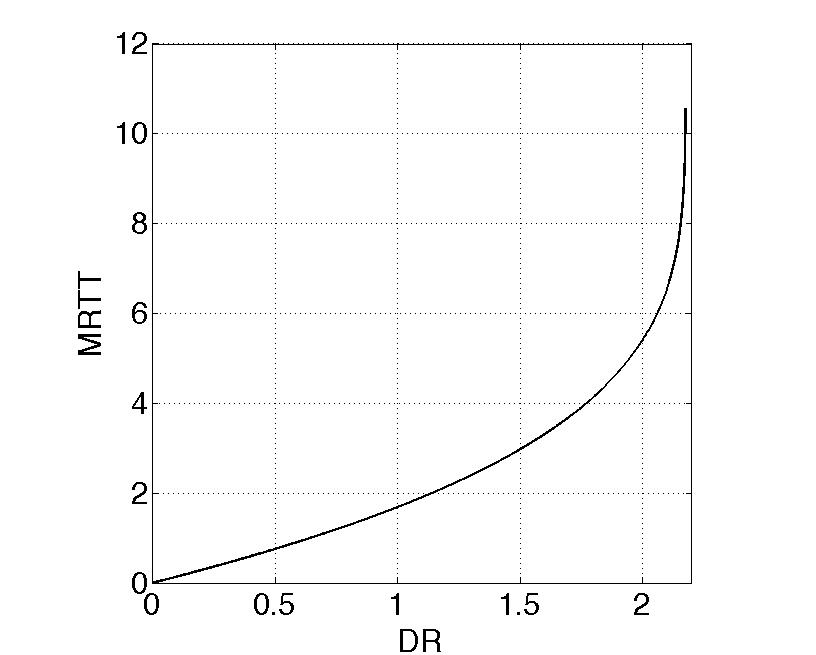}
\caption{\footnotesize \sl MRTT, in dB versus DR}\label{fig:reltt_versus_decodingratio}
\end{figure}

For a fixed $\overline{P_{r}}$ and $\tau_d$, energy is reduced for OLA-T by minimizing $\epsilon$ (and hence, $\tau_b$) . We next derive an expression for $\epsilon_{\textrm{min}}$ for a fixed DR. First,  we re-write (\ref{eq:difference_equation}) as shown below.
\begin{equation}
\mathbf{z_{k+1}}=\bf{A} \mathbf{z}_k,
\end{equation}
where $\mathbf{z_k}=\Big(r_{d,k}^{2},r_{b,k}^{2}\Big)^T$, and $ \mathbf{A} = \Big(\alpha(\tau_{d})+1,-\alpha(\tau_d); \alpha(\tau_{b})+1,-\alpha(\tau_{b})\Big)^T$. It can be seen that $A_1$ and $A_2$ are the eigenvalues of $\bf{A}$. For infinite network broadcast, the OLA rings must continue to grow implying that the system described by (\ref{eq:difference_equation}) must be ``unstable," i.e., $|A_1|>1$ \cite{elaydi}. Since, radii are always positive and $\alpha(\tau_{d})>\alpha(\tau_{b})>0$ by design,  $A_1>1$ becomes a necessary and sufficient condition for infinite network broadcast. Setting $A_1=1$ would give us an expression for the minimum $\epsilon$.
\begin{eqnarray*}
A_1 &=&1,\\
\Rightarrow \alpha(\tau_{d})-\alpha(\tau_{b})&=&1,\\
\Rightarrow \frac{1}{\biggl[\exp\left(\frac{\tau_{d}}{\overline{P_{r}}\pi}\right)-1\biggr]}-1 &=& \frac{1}{\biggl[\exp\left(\frac{\tau_{d}+\epsilon}{\overline{P_{r}}\pi}\right)-1\biggr]}.
\end{eqnarray*}
Collecting the $\tau_d$ terms and solving for $\epsilon$ results in 
\begin{equation}
\epsilon_{\textrm{min}} =(-1)\Bigg\{\overline{P_{r}}\pi\ln\bigg[2-\exp\left(\frac{\tau_{d}}{\overline{P_{r}}\pi}\right)\bigg]+ \tau_{d}\Bigg\},
\end{equation}
and the following necessary and sufficient condition:
\begin{equation}\label{eq:upperbound2}
2\geq \exp\left(\frac{\tau_{d}}{\overline{P_{r}}\pi}\right) +\exp\left(\frac{-\tau_{d}-\epsilon}{\overline{P_{r}}\pi}\right).
\end{equation}
%However, finding a  analytically is very difficult, so we have attempted to find it numerically. Specifically, we evaluated $DD_k$ for all $k$ up to a large value $K_{\textrm{max}}$. If $M<K_{\textrm{max}}$ ($M$ is from the DD criterion), we declare $\tau_b$ to be admissible.
We remark that when $\epsilon\rightarrow\infty$, OLA-T becomes Basic OLA, and (\ref{eq:upperbound2}) becomes the same condition (\ref{eq:successcriteria}) that was derived in
\cite{asympanal}. So for infinite network broadcast using OLA-T, $\epsilon>\epsilon_{\textrm{min}}$, or equivalently, $\tau_b>{\tau_b}_\textrm{min}$. Fig. \ref{fig:reltt_versus_decodingratio} shows Minimum Relative Transmission Threshold (MRTT), ${\tau_b}_\textrm{min}/\tau_d$, in dB, versus the DR. For example, for $\textrm{DR} = 1$, the minimum transmission threshold is about 1.8 dB higher than the decoding threshold. It can also be inferred that theoretically, it is possible for OLA-T to achieve infinite network broadcast with an infinitesimally small ${\tau_b}_\textrm{min}$ and DR. However, a very small MRTT may not be very effective if the precision in the estimate of the SNR is not good enough.

%We find that the MRTT is sensitive to $K_{\textrm{max}}$ for small values of the DR. If $K_{\textrm{max}}$ is increased, MRTT will drop slightly, because a large $k$ may be required before $DD_k$ becomes positive (i.e. before the curve in Fig. \ref{fig:ola_radii_versus_level} starts to turn upwards). Similarly, MRTT is again sensitive to $K_{\textrm{max}}$ as the decoding ratio approaches the maximum specified in (\ref{eq:successcriteria}) because it becomes unbounded. The part of the curve that we show changes imperceptively as $K_{\textrm{max}}$ increases up to 10,000.

\subsection{Energy Consumption for a Given Broadcast}
In this section, we compare the total radiated energy during a successful OLA-T broadcast to that of a successful Basic OLA broadcast. As $\tau_{b}\rightarrow\infty$ (or $\epsilon \rightarrow\infty$), the OLA-T rings grow in thickness until they become the same as the OLA decoding levels as in \cite{asympanal}. On the other hand, as $\tau_{b}\rightarrow\tau_{d}$, one would expect the transmitting strips to start \textit{thinning out}. In other words, the inner and outer radii for each OLA become close and the OLA areas decrease. If the DR is allowed to diminish (e.g. as $\overline{P_{r}}$ increases for a fixed $\tau_d$) as $\tau_{b} \rightarrow \tau_{d}$, successful broadcast can be maintained in a continuum network, even though OLAs become very thin.

If the transmit energy consumptions for Basic OLA and OLA-T are compared for the same $\overline{P_{r}}$, then it can be shown that OLA-T saves over 50\% of the energy consumed by Basic OLA \cite{sensorcomm}. However, Basic OLA can achieve successful broadcast  at a lower $\overline{P_{r}}$ according to (\ref{eq:successcriteria}).  Hence, we need to compare these two protocols for a fixed value of $\tau_{d}$ (i.e. data rate) such that each is in its minimum energy configuration.

The energy consumed by OLA-T in the first $L$ levels is mathematically expressed, in energy units, as
$\xi^{L}=\overline{P_{r}}T_{s}\displaystyle\sum_{k=1}^{L}{\pi(r_{d,k}^{2}-r_{b,k}^{2})}$, where $T_{s}$ is the length of the message in time units.  The Fraction of transmission Energy Saved (FES) for OLA-T relative to Basic OLA can be expressed as 
\begin{equation}\label{eq:fes}
\textrm{FES}= 1-\frac{\overline{P_{r}}_{(\textrm{OT})}\displaystyle\sum_{k=1}^{L}{\left(r_{d,k}^{2}-r_{b,k}^{2}\right)}}{\overline{P_{r}}_{(\textrm{O})} r_{d,L}^{2}},
\end{equation}
where $\overline{P_{r}}_{(\textrm{OT})}$ and $\overline{P_{r}}_{(\textrm{O})}$ are the lowest values of $\overline{P_{r}}$ that would guarantee successful broadcast using OLA-T and Basic OLA, respectively. If we multiply the numerator and denominator of the ratio by $1/\tau_d$, and substitute $\overline{P_{r}}_{(\textrm{O})}/\tau_d$ by its smallest value of $\pi\ln2$, we can re-write (\ref{eq:fes}) as
\begin{equation}\label{eq:fes2}
\textrm{FES}= 1-\frac{\pi\ln2\displaystyle\sum_{k=1}^{L}{\left(r_{d,k}^{2}-r_{b,k}^{2}\right)}}{\left(\tau_d/\overline{P_{r}}_{(\textrm{OT})}\right) r_{d,L}^{2}}.
\end{equation}

\begin{figure}[htp]
\centering
\includegraphics[width = 3.3in]{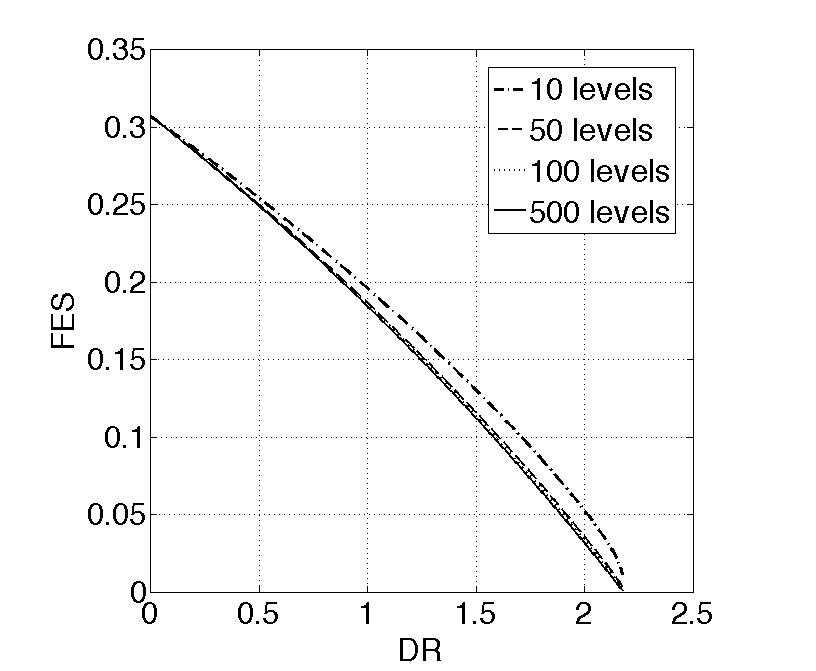}
\caption{\footnotesize \sl Variation of FES with DR and the numbers of levels (network size)}\label{fig:fes_versus_decodingratio_highlevels}
\end{figure}

Fig.\ref{fig:fes_versus_decodingratio_highlevels} shows FES versus DR for various network sizes (i.e. numbers of decoding levels or hops). We learn that there is a small dependence of FES on the number of levels, but it quickly diminishes after 50 levels. We observe that the FES is positive over the range of DR values we consider. For example, at $\textrm{DR} = 0.5$, FES is about 0.25. This means that at their respective lowest energy levels (OLA-T at ${\tau_b}_\textrm{min}$, and Basic OLA at $\overline{P_{r}}_{(\textrm{OT})}$), OLA-T saves about 25\% of the energy used by Basic OLA at this DR. FES increases as DR decreases and attains a maximum of about 32\% for infinitesimally small values of DR. 

The results given so far have been in terms of normalized units.  We would now like to consider some examples of un-normalized values for these variables to give an idea of what power levels and node densities can achieve the various values of DR and FES.  The DR ratio was previously defined as $\tau_{d}/\overline{P_{r}}$, where $\tau_{d}$ is the required SNR for decoding, $P_{r}$ is the normalized relay transmit power, and $\rho$ is the node density in number of nodes per area, where area is normalized by the square of the reference distance, $d_0^2$. Expanded in terms of un-normalized variables, we can write DR as
 \begin{equation}\label{eq:normalization_1}
\textrm{DR}=\frac{\bigg(\frac{\textrm{Receiver Sensitivity in mW}}{\sigma_n^2}\bigg)}{\bigg[\frac{P_tG_tG_r}{\sigma_n^2}\bigg(\frac{\lambda}{4\pi d_0}\bigg)^2\bigg]\bigg[\frac{(\#~\textrm{nodes})d_0^2}{\textrm{Area in}~\textrm{m}^2}\bigg]},
\end{equation}
where $P_t$  is the relay transmit power in mW,  $G_t$ and $G_r$ are the transmit and receive antenna gains,  $\sigma_n^2$ is the thermal noise power,  $\lambda$ is the wavelength in meters, and $d_0$  is the reference distance in meters. Suppose that the radio frequency is 2.4 GHz ($\lambda=0.125$ m), and the antennas are isotropic ($G_t=G_r=1$).  Then (\ref{eq:normalization_1}) can be simplified to
 \begin{equation}\label{eq:normalization_2}
\textrm{DR}=\frac{\bigg(\textrm{Receiver Sensitivity in mW}\bigg)10^4}{\big[P_t \textrm{~in mW}\big]\bigg[\frac{(\#~\textrm{nodes})}{\textrm{Area in}~\textrm{m}^2}\bigg]}.
\end{equation}

\begin{table}[!t]
\renewcommand{\arraystretch}{1.0}
\caption{Examples of un-normalized variables}
\label{normalization_jcn}
\centering
\begin{tabular}{|c|c|c|c|c|c|} 
\hline
Example & $P_t$  & Node Density & RX sens.& $d_{nn}$& DR\\
& (dBm) &(nodes/area) & (dBm) & (m)&\\\hline
1 & -56.00 & 2.65 nodes/$\textrm{m}^2$ & -90.00 & 0.61 & 1.5\\\hline
2 & -56.00 &2.65 nodes/$\textrm{m}^2$ & -94.77 & 0.61 & 0.5\\\hline
3 & -34.95 & 1 node/16~$\textrm{m}^2$ & -90.00 & 4.00 & 0.5\\\hline
4 & -43.98 & 1 node/4~$\textrm{m}^2$ & -90.00 & 2.00 & 1.0\\\hline
5 & -20.97 &9 nodes/3.60~$\textrm{km}^2$ & -90.00 & 20.00 & 0.5\\\hline
\end{tabular}
\end{table}

Table \ref{normalization_jcn} shows five different examples of un-normalized variables and their resulting $d_{nn}$ and DR values. We observe that $\textrm{DR} = 0.5$ can be obtained in Examples 2, 3, and 5, ranging from high density (2.65 nodes/$\textrm{m}^2$) to low density (9 nodes/3.60~$\textrm{km}^2$).  We also observe that the high density cases, Examples 1 and 2, correspond to very low transmit powers.

\section{OLA-T Broadcast with Variable $\epsilon$}\label{sec:olavt}
The OLA-based cooperative transmission techniques presented so far in the paper involve just a single \textit{fixed} $\epsilon$ for the whole wireless system. A shortcoming of this technique is that the radii growth is polynomial and the OLA rings keep growing bigger, expending more energy than is needed, to cover a given network area. We are motivated, therefore, to investigate how much more energy can be saved by letting each level have a different $\epsilon$.  We call the resulting broadcast protocol OLA-Variable Threshold (OLA-VT).

The Genetic Algorithm (GA) is adopted to determine the sequence of $\{\epsilon_{k}\}$ that yields the minimum OLA-T energy per broadcast, for a given $\tau_d$, $\overline{P_{r}}$, and fixed number of decoding levels. Two different constraints are considered. For each constraint, the radii are computed for the optimized $\{\epsilon_{k}\}$, and the FES is computed, assuming Basic OLA is in its minimum energy configuration. Figure \ref{fig:ola_radii_versus_level} suggests that a criterion for successful broadcast is eventual upward concavity of the curve. To capture this, we define the $k$-th Double Difference (DD) as $DD_{k}= (r_{d,k+2}-r_{d,k+1})-(r_{d,k+1}-r_{d,k})$.  Constraint Type 1 is that $DD_{k}>0$ for $k\geq4$; the total number of levels or hops is fixed, but no constraint is made on the physical size of the network. Constraint Type 2, on the other hand, fixes the number of levels \textit{and} the physical size of the network. The key difference is that Constraint Type 2 requires that the outer radius of the last decoding level be greater than the specified network radius. 

\begin{figure}[htp]
\centering
\includegraphics[width = 3.3in]{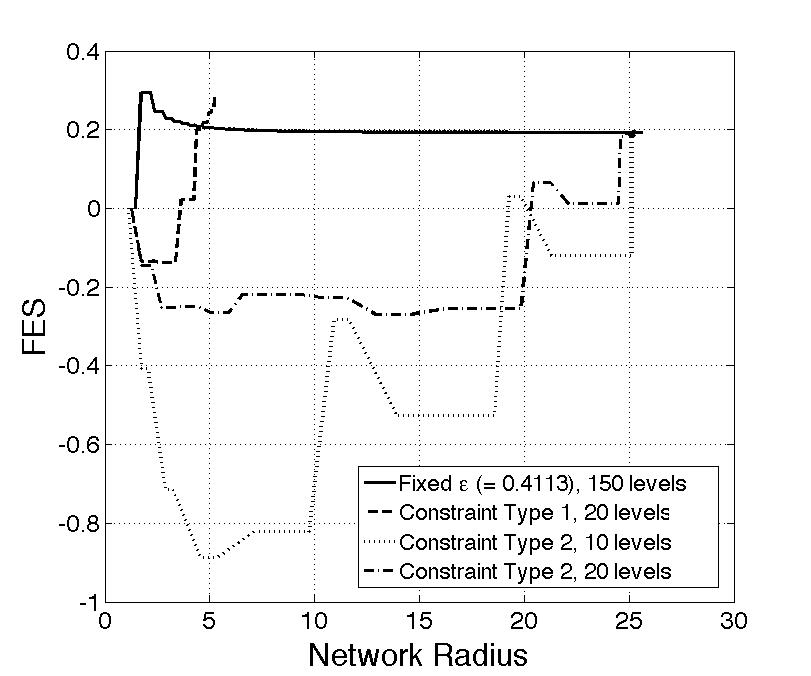}
\caption{\footnotesize \sl FES comparisons for variable $\epsilon_{k}$ versus fixed $\epsilon$}\label{fig:fes_olavt}
\end{figure}

Fig.\ref{fig:fes_olavt} plots the FES as a function of network radius. Constraint Type 1 is evaluated for a maximum of 20 levels (dashed line), and Constraint Type 2 is evaluated for 10 levels (dotted line) and 20 levels (dash-dot). Both Constraint Type 2 cases fix the network radius to be 25 distance units. As an example, for the 20-level case, the Constraint Type 2 algorithm minimizes broadcast energy with the constraint that $r_{d,20}>25$. The fixed $\epsilon$ case (solid line) is included for reference and requires 150 levels to reach a radius of 25. All OLA-T and -VT examples share the same DR of 0.9, and $P_s/\overline{P_{r}}$ of 4.31 dB. The fixed $\epsilon$ case uses the MRTT of 1.56 dB. The points on each curve are the FES values calculated for each radius in the sequence $\{r_{d,1}, r_{b,2}, r_{d,2}, r_{b,3},\ldots\}$. Since the FES is a function of whole levels and not partial levels, we just define the FES for $r_{b,k}$ to be equal to the FES for $r_{d,k-1}$; this enables us to identify OLA widths as the widths of the flat parts of the curve. The first non-zero point represents the FES at $r_{d,1}$, since the FES at $r_{b,1}$ is zero. Even though the constraints involve a fixed number of levels or physical network size, the FES value at a particular radius, $r$, indicates the FES as though the network were truncated to have radius $r$. For example, after 2 OLAs (i.e. at the right edge of the second plateau), the constant $\epsilon$ curve indicates an FES of about 0.25 at a radius of about 3. This means that a network of radius 3 that uses the fixed $\epsilon$ of 0.4113 to form two OLAs will achieve 25\% energy savings over the minimum energy Basic OLA for  network of radius 3.

We note that the Ònetwork radiusÓ in Fig.\ref{fig:fes_olavt} is normalized by the reference distance.  This means that if  $d_0=1$ m, then a network with normalized radius of 5 has an un-normalized radius of 5 m.  On the other hand, if $d_0=100$ m, then the same normalized network radius represents an un-normalized radius of 500 m.  When $d_0$  increases in (\ref{eq:normalization_1}), to maintain the same normalized relay transmit power, the un-normalized transmit power must increase by a factor of $d_0^2$, and to maintain the same normalized density, the un-normalized density must decrease by $d_0^2$.  In other words, if $d_0$ increases by a factor of 10, then the DR and hence the FES can be conserved by having the nodes spread out so that inter-node un-normalized distances increase by a factor of 10, and having the un-normalized relay transmit power increase by a factor of 100.

Let us first compare the Constraint Type 1 curve to the fixed $\epsilon$ curve.  We notice that the fixed $\epsilon$ curve starts high and then decays down to about 0.2.  The Constraint Type 1 curve, on the other hand drops to negative FES values and then climbs up to a final value of about 0.3.  That the final value of 0.3 is higher than the FES of the fixed $\epsilon$ curve for the same network radius of approximately 5 is evidence that variable  $\epsilon$ can be more energy efficient than fixed  $\epsilon$. The FES is negative because the $\overline{P_{r}}$ for OLA-T is larger than the $\overline{P_{r}}$ for Basic OLA, while the first few OLAs of OLA-T are allowed to be large and comparable to the first few OLAs of Basic OLA in size. The step sizes or hop distances for the fixed $\epsilon$ curve decrease smoothly with network radius, while the step sizes for the Constraint Type 1 curve are on the same order for the first 4 levels, until the FES reaches 0.2, and then the step sizes decrease significantly.  Relatively small step sizes should be OK as long as the density is high enough so that the OLA ring is several  $d_{nn}$ thick.

Constraint Type 2 curves drop down to much lower FES values and eventually climb back up to about 0.2.  At first glance, it may seem that the variable $\epsilon$ case does no better than the constant $\epsilon$ case, until one considers that the variable $\epsilon$ case reaches the same FES in only 10 or 20 steps, while constant $\epsilon$ requires 150 steps.  A  $d_0$ of 10 m, for example, would result in a Constraint Type 2 network of radius 250 m, with OLA sizes that would be reasonable for $\rho$ on the order of 1 node/4~$\textrm{m}^2$, as in Example 4 in Table \ref{normalization_jcn}.

\section{Simulations with Fading and Finite Node Densities}\label{sec:continuum_validity}
Throughout our analysis we have assumed the \textit{deterministic model} (nodes in the network relay in orthogonal channels) and a continuum of nodes (node density, $\rho, \rightarrow \infty$) \cite{asympanal}. In this section, we use Monte Carlo simulations to show that results based on the continuum and deterministic assumptions can be approximated by networks of finite density with Rayleigh fading channels.  First we will address the continuum assumption  and second we will address the orthogonality assumption. We will evaluate performance in terms of the Probability of Successful Broadcast (PSB), where a successful broadcast is when 99\% of the nodes in the network can decode the message. Normalized values are used in each case.

\begin{figure}[htp] \centering
\includegraphics[width = 3.3in]{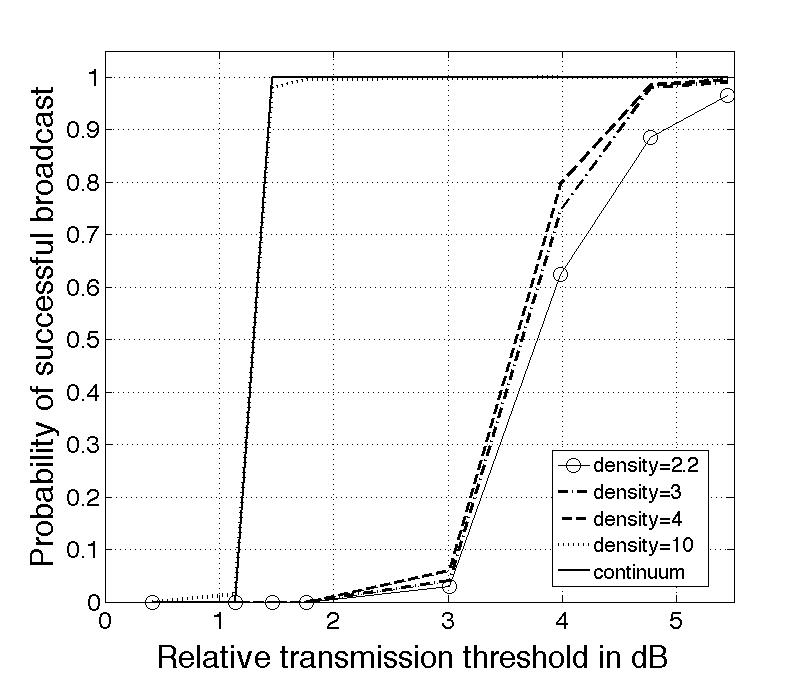}
\caption{\footnotesize \sl PSB as a function of RTT and node density, $\rho$}\label{fig:prob_succ_broad_versus_density}
\end{figure}

Fig. \ref{fig:prob_succ_broad_versus_density} is a plot of the PSB as a function of RTT in dB for various choices of network densities. The step function that represents the continuum assumption is also plotted. The Monte Carlo results have been obtained from a simulation of 400 random networks. The parameters used for the Monte Carlo trials are as follows. The nodes were uniformly and randomly distributed over a circular area of radius 17 distance units with the sink node located at the center of the network area. The source power, $P_s$, was chosen to be 3 and decoding threshold, $\tau_{d}$ was 1. Nodes in the first level used a fixed $\epsilon$ of 2.5 (in terms of RTT in dB, 5.44) for all the trials. RTT for all the other levels follow the horizontal axis. From Fig. \ref{fig:prob_succ_broad_versus_density}, we observe that at high values of network densities (10 in this case), the results from the trials approach the continuum plot, which represents an upper bound on the PSB for an RTT. At $d_0=1$ m, $\rho=10$ corresponds to an un-normalized node density of 10 nodes/$\textrm{m}^2$. At this high node density, the PSB drops sharply at a value of RTT, below which broadcast fails.

\begin{figure}[htp] \centering
\includegraphics[width = 3.3in]{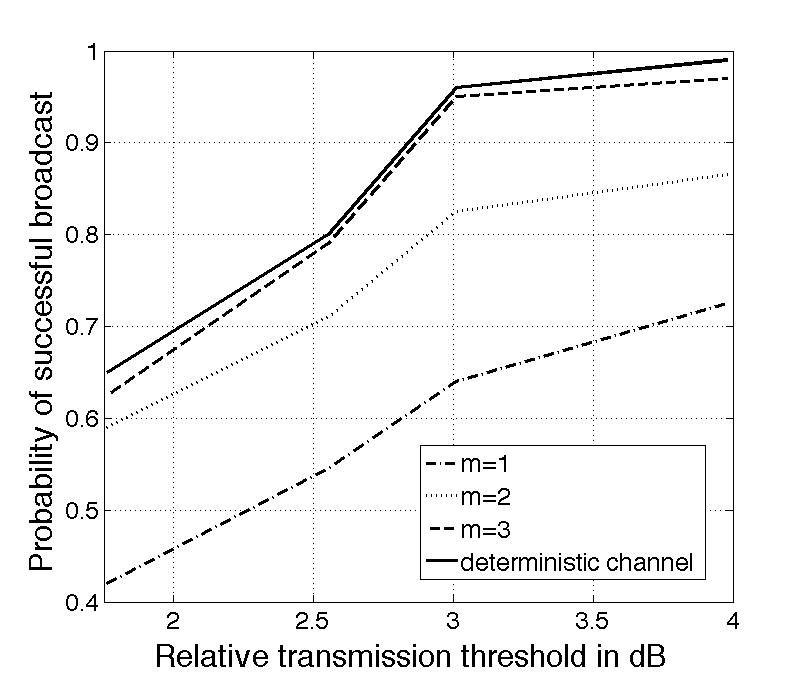}
\caption{\footnotesize \sl PSB as a function of RTT and diversity order, $m$}\label{fig:prob_succ_broad_versus_m}
\end{figure}

Next, we consider fading channels. This time, we assume 1500 nodes to be uniformly and randomly distributed over the same circular network area. All the other parameters are kept unchanged throughout the 100 network realizations. Rayleigh fading is assumed.  The transmitted signals are assumed to be Direct Sequence Spread Spectrum (DSSS), and the receivers are assumed to have a 4-th order RAKE processor. Let $m$ denote the desired diversity order. We assume that the chip width is larger than the delay spread of the real multipath channel.  To ensure diversity gain in the RAKE receiver, each relay chooses a transmit delay randomly from a window of 4 chip delay choices. $m=1$ means there is just one transmission path and is no diversity \cite{wei}, \cite{mudumbai}. Likewise, $m=2$ means 2 fingers are excited in the RAKE receiver. We assume  the chip pulse width to be 500 ns which is consistent with MICA sensor mote specifications \cite{xbow}. From Fig. \ref{fig:prob_succ_broad_versus_m}, we observe that with $m=3$, the performance in a fading channel is very similar to the deterministic channel.

\section{Conclusion}
In this paper, we proposed and analyzed a novel energy-efficient strategy that leverages the cooperative advantage in multi-hop wireless networks. By self-scheduling the transmissions in a network, significant energy savings are realized. We have shown how  OLA-T can save a maximum of about 32\% of the energy of a Basic OLA broadcast, and OLA-VT saves additional energy with no overhead and no central control. For fixed-size networks, OLA-VT simplifies boundary-matching. The trade-offs between the FES and the number of hops over a given network size were discussed using the OLA-VT. Further work is needed in terms of performance for practical estimators of SNR and practical synchronization.

\section*{Acknowledgment}
The authors are thankful for the reviewers' comments. The authors are also thankful to Prof. Shui-Nee Chow, from the School of Mathematics at the Georgia Institute of Technology for helpful discussions.

% that's all folks

\begin{thebibliography}{1}
\bibitem{sendonaris}A.~Sendonaris, E.~Erkip, and B.~Aazhang,``User Cooperation -- part i: System Description, part ii: Implmentation Aspects and Performance Analysis," \emph{IEEE Trans. Commun.}, vol.~51, no.~11, pp.~1927--48, Nov.~2003.

\bibitem{laneman}J.~N.~Laneman, D.~Tse, and G.~W.~Wornell,``Cooperative Diversitry in Wireless Networks: Efficient Protocols and Outage Behaviour,"\emph{IEEE Trans. Inf. Theory}, vol.~50, no.~12, pp.~3063--80, Dec.~2004.

\bibitem{hong}Y.~W.~Hong and A.~Scaglione,``Energy-efficient Broadcasting with Cooperative Transmissions in Wireless Sensor Networks," \emph{IEEE Trans. Wireless Comm.}, vol.~5, no.~10, pp.~2844--55.

\bibitem{Ni} Sze-Yao Ni, Yu-Chee Tseng, Yuh-Shyan Chen, and Jang-Ping Sheu, ``The Broadcast Storm Problem in a Mobile Ad hoc Network," \emph{Proc. ACM/IEEE International Conf. on Mobile Comput. and Netw.}, 1999, pp.~151--62.

%\bibitem{prim} R.~C.~Prim, ``Shortest Connection Networks and Some Generalizations," {\emph Bell System Technical Journal}, vol.~36, pp.~1389--1401, 1957.

\bibitem{wieselthier} J.~Wieselthier, G. Nguyen, and A. Ephremides, ``On the Construction of Energy-Efficient Broadcast and Multicast Trees in Wireless Networks," \emph{Proc. IEEE INFOCOM}, Mar.~2000, pp.~585--594.

\bibitem{cartigny}J.~Cartigny and D.~Simplot,``Border Node Retransmission based Probabilistic Broadcast Protocols in Ad Hoc Networks," \emph{Proc. HICSS},  2003.

\bibitem{cartigny1}J.~Cartigny, F.~Ingelrest, and D.~Simplot,``RNG Relay Subset Flooding Protocol in Mobile Ad Hoc Networks," \emph{International J. Foundations of Comput. Sci.}, vol.~14, no.~2, pp.~253--265, Apr. ~2003.

\bibitem{ola}A.~Scaglione, and Y.~W.~Hong,``Opportunistic Large Arrays: Cooperative Transmission in Wireless Multi-hop Ad Hoc Networks to Reach Far Distances," \emph{IEEE Trans. on Signal Process.}, vol.~51, no.~8, pp.~2082--92, Aug.~2003.

\bibitem{wei} S.~Wei, D.~L.~Goeckel, and M.~Valenti, ``Asynchronous Cooperative Diversity," \emph{Proc. CISS}, Mar.~2004.

%\bibitem{savvides}A.~Savvides, C.~C.~Han, and M.~B.~Srivastava,``Dynamic Fine-Grained Localization in Ad Hoc Networks of Sensors,"\emph{Proc. 7th Annual International Conference on Mobile Computing and Networking}, pp.~166--179, Rome, Italy, 2001.

%
%\bibitem{sundaram}N.~Sundaram, P.~Ramanathan,``Connectivity based Location Estimation Scheme for Wireless Ad Hoc Networks,"\emph{Proc. IEEE GLOBECOM}, vol.~1, pp.~143--147, Nov.~2002.

\bibitem{ivana}I.~Maric and R.~D.~Yates,``Cooperative Multi-Hop Broadcast for Wireless Networks," \emph{IEEE J. Sel. Areas Commun.}, vol.~22, no.~6, pp.~1080--88, Aug.~2004.

\bibitem{ahluwalia} A.~Ahluwalia, E.~Modiano, and L.~Shu,``On the Complexity and Distributed Construction of Energy-Efficient Broadcast Trees in Static and Ad Hoc Wireless Networks," \emph{Proc. CISS}, Mar.~2002, pp.~193--202.

\bibitem{liang}W.~Liang,``Constructing Minimum-Energy Broadcast Trees in Wireless Ad hoc Networks," \emph{Proc. MOBIHOC}, Jun.~2002, pp.~112--22.

\bibitem{optimal}B.~Sirkeci-Mergen and A.~Scaglione,``On the Optimal Power Allocation for Broadcasting in Dense Wireless Networks,"\emph{Proc. IEEE ISIT}, 2006.

\bibitem{power}B.~Sirkeci-Mergen and A.~Scaglione,``On the Power Efficiency of Cooperative Broadcast in Dense Wireless Networks,"\emph{IEEE J. Sel. Areas Commun.}, vol.~25, no.~2, pp.~497--507, Feb.~2007.

\bibitem{asympanal}B.~Sirkeci-Mergen, A.~Scaglione, G.~Mergen,``Asymptotic analysis of multi-stage cooperative Broadcast in Wireless Networks," \emph{Joint special issue of the IEEE Trans. Inf. Theory and IEEE/ACM Trans. Net.}, vol.~52, no.~6, pp.~2531--50, Jun.~2006.

%\bibitem{continuum}B.~Sirkeci-Mergen and A.~Scaglione,``A Continuum Approach to Dense Wireless Networks with Cooperation," \emph{Proc. IEEE INFOCOM 2005}, Vol.~4, pp.~2755--63, 2005.

\bibitem{sensorcomm}L.~Thanayankizil, A.~Kailas, and M.~A.~Ingram,``Energy-Efficient Strategies for Cooperative Communications in Wireless Sensor Networks,"\emph{Proc. SENSORCOMM}, 2007.

%\bibitem{milcom}A.~Kailas, L.~Thanayankizil, and M.~A.~Ingram, ``Power Allocation and Self-Scheduling for Cooperative Transmission Using Opportunistic Large Arrays,"\emph{Proc., MILCOM}, 2007.

\bibitem{biagioni} E.~Biagioni, ``Algorithms for Communication in Wireless Multi-Hop Ad Hoc Networks Using Broadcasts in Opportunistic Large Arrays (OLA)," \emph{Proc. ICCCN}, Aug.~07, pp.~1111--16. 

\bibitem{globecom08} A.~Kailas and M.~A.~Ingram, ``Alternating Cooperative Transmission for Energy-Efficient Broadcasting," \emph{Submitted to GLOBECOM}, 2008.

\bibitem{elaydi} S.~Elaydi, ``Differential Equations: Stability and Control," CRC Press, 1991.

\bibitem{mudumbai} R.~Mudumbai, G.~Barriac, and U.~Madhow, ``Spread-spectrum Techniques for Distributed Space-Time Communication in Sensor Networks,Ó \emph{Proc. Asilomar Conf.}, vol.1, pp.~908--12, Nov.~2004. 

\bibitem{xbow} \texttt{www.xbow.com/Products/}

\end{thebibliography}
\end{document}